\documentclass[twocolumn,showpacs,preprintnumbers,nofootinbib,groupedaddress,floatfix]{revtex4-1}
\usepackage{graphicx,amsmath,amssymb,xcolor,hyperref}

\usepackage{lipsum}
\usepackage{bm}
\usepackage{amsmath}
\usepackage{amssymb}
\usepackage{bbold}
\usepackage{graphicx}
\usepackage{svg}
\usepackage{comment}
\usepackage{natbib}

\usepackage[compat=1.1.0]{tikz-feynman}
\begin{document}
\title{Restoring Naturalness via Conjugate Fermions}
\author{Andrei Angelescu}
\email{andrei.angelescu@mpi-hd.mpg.de}
\author{Andreas Bally}
\email{andreas.bally@mpi-hd.mpg.de}
\author{Florian Goertz}
\email{florian.goertz@mpi-hd.mpg.de}
\author{Maya Hager}
\email{maya.hager@mpi-hd.mpg.de}
\affiliation{Max-Planck-Institut f{\"u}r Kernphysik, Saupfercheckweg 1, 69117 Heidelberg, Germany}


\begin{abstract}
We propose a novel mechanism for cancelling the leading order contribution to the potential in composite Higgs scenarios. The mechanism relies on the splitting of a pseudoreal representation of the global symmetry into a complex representation and its conjugate of the unbroken group. We identify two cosets one of which includes a custodial symmetry. A numerical analysis is performed in a phenomenological three--site model and the resulting fine--tuning is analysed. The cancelling of the leading order potential results in a drastic reduction of the fine--tuning. For a symmetry breaking scale of the strong sector as high as $f=1600$ GeV, fine--tuning can be as good as $10\%$ or even better. We discuss a possible interpretation in the 5D holographic dual. Unique signatures of the model include quarks with baryon number $B=2/3$ with highly distinctive decays which can be looked for at the LHC. 
\end{abstract}
\maketitle
\textbf{Introduction.}
The origin of the weak scale, parametrised by the Higgs field, remains unknown. Due to its quadratic sensitivity to UV scales, an unnaturally large tuning seems necessary to separate the weak scale from, for instance, the Planck scale. Understanding if and how the solution to this hierarchy problem (HP) can be found at the CERN LHC is one of the most important objectives of high energy physics.

The HP can be elegantly addressed in the framework of Composite Higgs (CH)~\cite{Kaplan:1983fs,Kaplan:1983sm,Dugan:1984hq}, in which the Higgs is not an elementary scalar anymore. Quadratically sensitive loop corrections are tamed above the scale of compositeness, $m_\ast$, at which the Higgs resolves into its more fundamental constituents. In such models, the weak scale is no longer an input but dynamically generated by a new force which condenses at a symmetry breaking scale $f$, therefore producing the Higgs as a pseudo--Nambu--Goldstone boson (pNGB) of a spontaneous breaking of a flavor symmetry of this new sector, $G/H$, akin to the pions of QCD (see \cite{Contino:2010rs,Bellazzini:2014yua,Panico:2015jxa} for reviews). However, signs of compositeness are lacking at the LHC, requiring the symmetry breaking scale to reside at larger values, which increases the generic fine--tuning of CH models. In particular, light top partners at the symmetry breaking scale, necessary to generate a large mass for the top quark, are an ubiquitous prediction of CH models and a driving force behind the increased fine--tuning~\cite{Contino:2006qr,Csaki:2008zd,DeCurtis:2011yx,Matsedonskyi:2012ym,Panico:2012uw,Marzocca:2012zn,Pomarol:2012qf,Carmona:2014iwa,Goertz:2018dyw}. This has resulted in model--building addressing the anomalously light top partners~\cite{Panico:2012uw,Carmona:2014iwa,Blasi:2019jqc,Blasi:2020ktl,Blasi:2022hgi} and the tuning~\cite{Csaki:2017cep,Csaki:2018zzf,Blasi:2020ktl,Blasi:2022hgi,Cheng:2020dum,Durieux:2021riy} (see also~\cite{Bally:2022naz,Bally:2023lji,Chung:2023iwj} for top Yukawa--based solutions). Moreover, the situation is significantly worsened in generic CH due to the feature that the radiative Higgs potential generates the quartic interaction at subleading order with respect to the quadratic, the so--called double--tuning problem \cite{Panico:2012uw}. 

In this paper, we propose a novel mechanism that reduces the tuning by cancelling the leading order contribution to the Higgs potential, so that both quadratic and quartic arise only at fourth order in couplings. We show that the tuning is even less than the conventional minimal estimate and furthermore explain why the lightest composite resonances have not yet been observed. 

\textbf{Mirror Fermions.}
{\it The quadratic contribution of a chiral fermion $\psi$ to the pNGB potential of a coset $G/H$ is cancelled when a new chiral fermion $\psi'$ with conjugated gauge quantum numbers is added, called mirror fermion, if the fermions talk to the same composite operator in a pseudoreal representation $\bf{R}$ of the group $G$ which decomposes as $\bf{R} \rightarrow \bf{C} \oplus \Bar{\bf{C}}$ under $H$, with $\bf{C}$ a complex representation and $\Bar{\bf{C}}$ its complex conjugate.} 

The statement is proven by considering the general coset $G/H$. In addition to the spontaneous breaking $G/H$, $G$ is explicitly broken by partial compositeness \cite{Kaplan:1991dc,Contino:2003ve,Agashe:2004rs} of the Standard Model (SM) fields -- a linear mixing of strength $\lambda$ between elementary fields $\psi$ and composite operators $\mathcal{O}^{\bf{R}}$
\begin{equation}
    \mathcal{L}_{\textrm{PC}} = \lambda\, \Bar{\psi} \Delta \,\mathcal{O}^{\bf{R}} + \lambda^\prime\, \Bar{\psi}^\prime \Delta^\prime \,\mathcal{O}^{\bf{R}} + \text{ h.c}\,.
    \label{equ:partialCompositenessGeneral}
\end{equation}
Here, the spurion $
\Delta^{(\prime)}$ parametrises the incomplete embedding of $\psi^{(\prime)}$ in $G$ (where $\psi^{(\prime)}$ has been filled to form a full $G$ representation).
We employ the Callan–Coleman–Wess–Zumino mechanism (\cite{Coleman:1969sm,*Callan:1969sn}) to calculate the contribution to the Goldstone potential. The spurions are dressed as $\Delta^{(\prime)} U$ with the Goldstone Matrix $U$, which is the exponential of the broken generators $T^{\hat{a}}$ of $G$, each of which corresponds to a NGB degree of freedom $\Pi_{\hat{a}}$ (see, for example, \cite{Panico:2015jxa}) 
\begin{equation}
    U = \exp{\left(i \Pi_{\hat{a}} T^{\hat{a}}\right)}.
\end{equation}
After dressing, the spurions decompose under $H$ representations $\bf{C}$ and $\bar{\bf{C}}$ as $\Delta U = (\Delta^{\bf{C}}_D, \Delta^{\Bar{\bf{C}}}_D)$ (analogously for $\Delta^\prime$), with $D$ differentiating the dressed spurions  from their undressed counterparts.

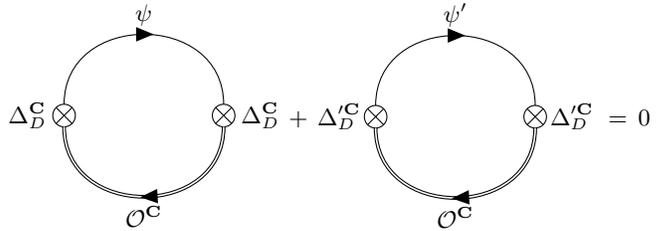
\begin{figure}
\centering
\[
\begin{aligned}
\begin{tikzpicture}
  \begin{feynman}
    \vertex [crossed dot] (i1) {\( \)} ;
    \vertex [above right =of i1] (j1) {\( \)};
    \vertex [below right =of j1,crossed dot] (i2) {\( \)};

    \diagram* {
       (i1) -- [fermion, half left, edge label=\(\psi\)] (i2) -- [double, half left, with arrow=1.5cm,edge label=\(\mathcal{O}^{\bf{C}}\)] (i1)
    };
    \vertex [right=1em of i2] {\(\quad\Delta^{\bf{C}}_D\)};
    \vertex [left=1em of i1] {\(\Delta^{\bf{C}}_D\quad\)};
  \end{feynman}
\end{tikzpicture}
\end{aligned}
+
\begin{aligned}
\begin{tikzpicture}
  \begin{feynman}
    \vertex [crossed dot] (i1) {\( \)} ;
    \vertex [above right =of i1] (j1) {\( \)};
    \vertex [below right =of j1,crossed dot] (i2) {\( \)};
    \diagram* {
       (i1) -- [fermion, half left,edge label=\(\psi^\prime\)] (i2) -- [double, half left, with arrow=1.5cm,edge label=\(\mathcal{O}^{\bf{C}}\)] (i1)
    };
    \vertex [right=1em of i2] {\(\quad\Delta'^{\bf{C}}_D\)};
    \vertex [left=1em of i1] {\(\Delta'^{\bf{C}}_D\quad\)};
  \end{feynman}
\end{tikzpicture}
\end{aligned}
=\,0
\]
\caption{Cancellation mechanism of the quadratic contribution in terms of Feynman diagrams.}
\label{fig:FeynmanCancellation}
\end{figure}

We note that, after contracting the $H$ indices, the product of spurions parametrising the embedding of the field $\psi$ in the representation $\bar C$ is the same as the one parametrising the embedding of the conjugate field $\psi^\prime$ in the conjugate representation $C$ (see Appendix \ref{app1} for the mathematical proof), i.e.
\begin{equation}
   {\rm Tr} \left[ (\Delta^{\Bar{\bf{C}}}_D)^\dagger \Delta^{\Bar{\bf{C}}}_D \right] = 
   {\rm Tr} \left[ (\Delta'^{\bf{C}}_D)^\dagger \Delta'^{\bf{C}}_D \right]
    \label{equ:trafoInvariants2}.
\end{equation}
Now, we show how the lowest order Feynman diagrams that contribute to the potential cancel. These diagrams can be seen in Fig.~\ref{fig:FeynmanCancellation} where in the first (second) diagram a $\psi$ ($\psi^\prime$) runs in the loop. Moreover, the loop is closed on the composite side via either a $\bf{C}$ or $\Bar{\bf{C}}$ two--point function. We will only show the cancellation for the $\bf{C}$ diagram, the $\Bar{\bf{C}}$ diagram follows analogously. Summing both diagrams gives the following two spurion combinations which contain the pNGB dependence
\begin{equation}
    V^{\bf{C}} \propto\lambda^2 {\rm Tr} \left[ (\Delta^{\bf{C}}_D)^\dagger \Delta^{\bf{C}}_D \right] + \lambda'^2 {\rm Tr} \left[ (\Delta'^{\bf{C}}_D)^\dagger \Delta'^{\bf{C}}_D \right],
\end{equation}
whereas the dependence on the composite sector factorises out. Using Eq.~\eqref{equ:trafoInvariants2} one can rewrite the above expression as 
\begin{equation}
    V^{\bf{C}} \propto\lambda^2 {\rm Tr} \left[ (\Delta^{\bf{C}}_D)^\dagger \Delta^{\bf{C}}_D \right] + \lambda'^2 {\rm Tr} \left[ (\Delta^{\Bar{\bf{C}}}_D)^\dagger \Delta^{\Bar{\bf{C}}}_D \right].
\end{equation}
If $\lambda=\lambda^\prime$, which follows for the minimal realization where $\psi$ and $\psi^\prime$ are part of the same incomplete $G$ representation, the pNGB dependence drops out due to the unitarity of the Goldstone matrix
\begin{equation}
    V^{\bf{C}} \propto \lambda^2 {\rm Tr} \left[ \Delta^\dagger U U^\dagger \Delta \right] = \lambda^2 N
    \label{eq:vacEnergyContr}
\end{equation}
and we are left with a contribution to the vacuum energy proportional to the fermionic degrees of freedom N. We emphasise that the origin of this cancellation mechanism is inherent in the decomposition of the pseudoreal representation $\bf{R}=\bf{C}\oplus \bar{\bf{C}}$ into a complex and its conjugate under the unbroken group $H$.

\begin{figure}
\centering
\[
\begin{aligned}
\begin{tikzpicture}
  \begin{feynman}
    \vertex [crossed dot] (i1) {\( \)} ;
    \vertex [above right =of i1,crossed dot] (i3) {\( \)};
    \vertex [below right =of i1,crossed dot] (i4) {\( \)};
    \vertex [below right =of i3,crossed dot] (i2) {\( \)};

    \diagram* {
       (i1) -- [fermion, quarter left,edge label=\(\psi'\)] (i3) -- [fermion, quarter left,edge label=\(\Tilde{\psi}'\)] (i2) --[fermion, quarter left,edge label=\(\psi'\)] (i4) -- [double, quarter left, with arrow=1cm,edge label=\(\mathcal{O}^{\bf{C}}\)] (i1)
    };
    \vertex [right=1em of i2] {\(\quad \quad \, m_E \,\)};
    \vertex [left=1em of i1] {\(\Delta'^{\bf{C}}_D\quad\)};
    \vertex [below=1.1em of i4] {\(\Delta'^{\bf{C}}_D\)};
    \vertex [above=1.1em of i3] {\(m_E\, \)};
  \end{feynman}
\end{tikzpicture}
\end{aligned}
= \, \mathcal{O}\Big(\big(\frac{m_E}{m_*} \lambda' \big)^2\Big)
\]
\caption{Remaining quadratic contribution in the presence of a Dirac mass for the mirror fermion.}
\label{fig:FeynmanCancellationRealistic}
\end{figure}
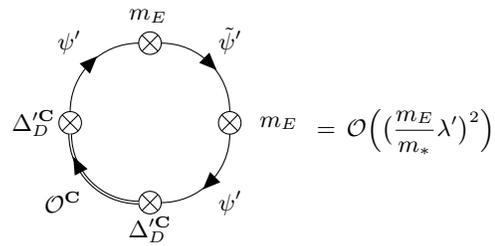

In realistic scenarios it is clearly not feasible to add a new massless chiral fermion. Therefore, it becomes necessary to introduce the opposite--chirality fermion $\Tilde{\psi}'$ and an elementary Dirac mass $m_E$ between them. We assume that the opposite--chirality fermion does not talk to the composite sector, since otherwise there could be additional quadratic contributions to the Higgs potential. Still, a quadratic contribution remains, which is however suppressed by $\left(\frac{m_E^2}{m_*^2}\right)$ (see Fig. \ref{fig:FeynmanCancellationRealistic}), where $m_*= g_* f \gg f$ is the resonance scale, with $g_*$ the coupling in the strong sector. 


\textbf{Holographic Completion.}
Although a light Dirac mass for the mirror fermion is technically natural and thus no hierarchy problem is introduced by its presence, it does beg the question: What should be its natural scale? The cancellation mechanism requires $m_E\lesssim m_*$ which could introduce a $\textit{coincidence}$ problem. It turns out $m_E$ has a very elegant origin once possible UV completions for the above model are considered. In the holographic dual of these models, inspired by the AdS/CFT correspondence~\cite{Arkani-Hamed:2000ijo,Rattazzi:2000hs}, where the pNGBs arise as the fifth component of a five dimensional gauge field in warped space with a UV/IR brane at $z=R/R^\prime$~\cite{Contino:2003ve,Agashe:2004rs}, the partial compositeness hypothesis from Eq.(\ref{equ:partialCompositenessGeneral}) is equivalent to embedding the elementary fermions within 5D bulk fermions transforming under $\bf{R}$~\cite{Contino:2004vy}. In contrast, the opposite--chirality fermion $\Tilde{\psi}'$ does not talk to the composite sector and therefore correspond to a UV brane--localised fermion in the holographic dual. Then, the Dirac mass corresponds to a UV brane--localised mass mixing between the brane fermion $\Tilde{\psi}'$ and the bulk fermion $\psi'$:
\begin{equation}
    \int \textrm{d}^4 x \frac{M_{UV}}{\sqrt{R}}\bar{\Tilde{\psi}}'(x) \psi'(x,z=R) +\textrm{ h.c.},
\end{equation}
with $M_{UV}\sim \mathcal{O}(1)$. Due to the 5D nature of $\psi'(x,z)$, the resulting 4D mass depends on its localisation along the extra dimension, which is commonly parametrised by its dimensionless 5D mass $m\equiv c/R$. We find two regimes depending on whether the bulk fermion is UV--localised ($c>0.5$) or IR--localised ($c<0.5$):
\begin{equation}\label{eq:5Drel}
m_{E}\sim \frac{M_{\textrm{UV}}}{R} \times
    \begin{cases}
    1 \quad &(c>0.5)\\
    (R^\prime/R)^{c-1/2} \, (1-c) \quad &(c<0.5)
    \end{cases},
\end{equation}
and we see that for an IR--localised bulk fermion, one can recover exponentially smaller masses than the natural expectation of $\sim 1/R$ for UV masses. Furthermore, we expect an IR--localisation since the heavy SM fields are IR--localised, and it is the contribution of those to the Higgs potential that we wish to cancel\footnote{The exponential sensitivity of $m_E$ to $c$ is identical to the sensitivity of the SM fermion masses to their respective 5D localization parameters (or equivalently in partial compositeness, the sensitivity to the dimension of the composite operators) and therefore doesn't introduce a hierarchy problem. Instead, it is consistent with the resolution of the flavor hierarchy problem in 5D.}.

In 5D one also obtains naturally the equality $\lambda=\lambda^\prime$, by embedding the SM fermion and its mirror within the same bulk fermion of $G$, in agreement with the 4D interpretation.

\textbf{Concrete Models.}
Two minimal cosets fulfilling the above criteria include the color gauge group $SU(3)_c$ as part of the flavor symmetry $G$. These models, motivated by charge quantisation as in 4D Grand Unified Theories~\cite{Pati:1973uk,Georgi:1974sy} (GUTs), are known as composite GUTs~\cite{Frigerio:2011zg,Barnard:2014tla,Agashe:2005vg} or their 5D warped duals of gauge--Higgs grand unification~\cite{Hosotani:2015hoa,Furui:2016owe,Funatsu:2019xwr,Funatsu:2020znj,Lim:2007jv,Angelescu:2021nbp,Angelescu:2022obm,Angelescu:2021qbr}. Interestingly, these models predict extra colored pNGBs. The non--custodial coset $SU(6)/SU(5)$ provides a minimal realisation, with the pseudoreal representation
\begin{equation}
    \bf{20} \rightarrow \bf{10} \oplus \bar{\bf{10}}
\end{equation}
where the decomposition of the $\bf{10}$ of $SU(5)$ under the SM gauge group is
\begin{equation}
    \bf{10} \rightarrow {\bf (3,2)}_{1/6} \oplus {\bf (3^*,1)}_{-2/3} \oplus  {\bf (1,1)}_1\,.
\end{equation}
However, the model is constrained by large corrections to the $T$ parameter. Instead, the custodial coset $SO(11)/SO(10)$, with the pseudoreal representation
\begin{equation}
    \bf{32} \rightarrow \bf{16} \oplus \bar{\bf{16}},
\end{equation}
also satisfies the criteria but does not generate a T parameter at tree--level.

Since the biggest source of explicit $G$--breaking stems from the top quark, we can focus on the right-handed top singlet $t_R$ and the left--handed quark doublet $q_L$ in the following analysis, whose contribution will be cancelled by two mirror fermions $\omega_R$ and $\theta_L$ respectively. As the $\bf{16}$ contains a $\bf{10}$ of $SU(5)$, all four fields fit into the $\bf{20}$ of $SU(6)$ or the $\bf{32}$ of $SO(11)$. The linear mixing strength $\lambda_{R/L}$ in the IR is expected to depend on the scaling dimension $d_{L/R}$ of the composite operator $\mathcal{O}^{\bf{R}}_{L/R}$ and its UV value $\lambda_{\text{UV}}$ \cite{Contino:2003ve,Agashe:2004rs,Contino:2004vy,Contino:2010rs,Panico:2015jxa}
\begin{equation}
    (\lambda_{\text{IR}})_{R/L} \sim (\lambda_{\text{UV}})_{R/L} \left( \frac{\Lambda_{\text{IR}}}{\Lambda_{\text{UV}}} \right)^{d_{L/R}- 5/2}.
\end{equation}
As discussed above, in the minimal representation where the SM field and its mirror fermion are embedded in the same incomplete representation of $G$, the strength of the partial compositeness interaction for both fermions is identical and the Lagrangian becomes
\begin{eqnarray}\label{PChypo}
    \mathcal{L}_{\text{PC}} = && \lambda_R \left( \Bar{t}_R \Delta^{t_R} + \Bar{\omega}_R \Delta^{\omega_R} \right) \mathcal{O}^{\bf{R}}_L + \text{h.c.} \, \nonumber \\
    && + \, \lambda_L \left( \Bar{q}_L \Delta^{q_L} +  \Bar{\theta}_L \Delta^{\theta_L} \right) \mathcal{O}^{\bf{R}}_R + \text{h.c.} \nonumber \\
    && + \, m_{\omega} \, \Bar{\omega} \omega + m_{\theta} \, \Bar{\theta} \theta,
\end{eqnarray}
including the Dirac masses for the mirror fermions. 

Gauge boson contributions will be neglected as they are subleading. Moreover, the numerical analysis in the three--site model in the next section for the fermion sector is identical for both considered cosets, fully determined by the symmetry properties of the pseudoreal representation, namely $\bf{R}\rightarrow \bf{C}\oplus \bar{\bf{C}}$.

For a complete modelling of the third generation of quarks\footnote{The lighter two generations can be embedded identically, although the cancellation effect would not hold as the resulting vector mass for the exotic fermion Eq.\eqref{eq:5Drel} would be unsuppressed due to the UV localization of the lighter generation. However, since the lighter generation contributions to the Higgs potential are negligible, this does not affect the numerical results.}, one must include the right--handed bottom quark $b_R$ in the partial compositeness Lagrangian with an associated composite operator $\mathcal{O}_L^{\bf{R'}}$ in a representation $\bf{R^\prime}$. Although its contribution to the Higgs potential is negligible due to the small bottom mass, the associated composite operator will mix with the ones of the top and exotic sector therefore impacting their mass spectrum. In order for the $b_R$ to connect to the $q_L$, the representation should decompose as $\bf{R^\prime} \rightarrow \bf{C} \oplus ...$ under $H$. Once $G$ is spontaneously broken, the composite operators $\mathcal{O}^{\bf{R}}_R$ and $\mathcal{O}^{\bf{R^\prime}}_{L}$ mix and induce a mass for the bottom quark. 

For the coset $SO(11)/SO(10)$, the minimal choice for  $\bf{R^\prime}$ is another $\bf{32}$. If we attempted to use the same $\bf{32}$ for the $b_R$ and $t_R$, there would be a degeneracy in their masses. For the $SU(6)/SU(5)$ the minimal option is a $\bf{15}$ which decomposes as $\bf{10}\oplus \bf{5}$ under $SU(5)$. 

As mentioned, the cosets contain more broken generators besides the ones of the Higgs doublet $\bf{(1,2)}_{1/2}$, and there will be more pNGBs. Both scenarios predict a scalar leptoquark $\bf{(3,1)}_{-1/3}$, and in $SU(6)/SU(5)$ there is an additional scalar singlet, whose generator corresponds to an unbroken global symmetry. Therefore, it remains massless unless the symmetry is broken by a different mechanism, i.e. by introducing a Majorana neutrino sector. For the leptoquark potential the cancellation in the fermion sector proceeds identically, and neither $q_L$ nor $t_R$ generate the potential at leading order.
However, the gauging of the strong sector is a large source for the potential of the leptoquark. Using NDA~\cite{Giudice:2007fh}, the leading potential can be estimated as~\cite{Angelescu:2022obm}
\begin{equation}
    V(S) \approx m_*^4 \frac{3\times 5}{64\pi^2}\frac{g_s^2}{g_*^2}\sin^2\left(\frac{\sqrt{2 S^\dagger S}}{f}\right).
\end{equation} 
The resulting mass for the leptoquark is then $m_S= (15\alpha_s/8\pi)^{1/2} m_*\approx (0.25 m_*)$. 

\textbf{Numerical Analysis.}
We proceed with a numerical analysis of the above setup in a multi--site model~\cite{Panico:2011pw,Contino:2006nn}. These phenomenological models are inspired by dimensional deconstruction~\cite{Arkani-Hamed:2001kyx} and by 5D models~\cite{Contino:2003ve}, retaining the useful features of finiteness of the Higgs potential while being computationally easier.

We will work in the three--site model, in which the first site models the elementary sector while the other two sites represent the composite sector. It is the lowest site--model in which the Higgs potential is fully calculable. The Higgs potential is determined with the Coleman--Weinberg formula~\cite{Coleman:1973jx}
\begin{equation}
    V_i (h) = - \frac{2 N_c}{8 \pi^2} \int \textrm{d}p p^3 \log \left( \det \left( M_i^{\dagger}(h) M_i(h) + p^2 \mathbb{1} \right) \right)
\end{equation}
where $N_c=3$, and $M_i$, with $i=T,E$, the mass matrices for the top and exotic sector respectively. As exotics we denote all mass eigenstates arising through the mirror fermion sector.

The numerical analysis is performed by scanning the composite masses over a range of $[-5f,5f]$, for a symmetry breaking scale of $f=1600$\,GeV. We assume $\lambda_L = \lambda_R \equiv \lambda$ for simplicity and match to the correct top mass $m_t(f)\approx 150$\,GeV.

\begin{figure}
    \centering
    \includegraphics[width=0.45\textwidth]{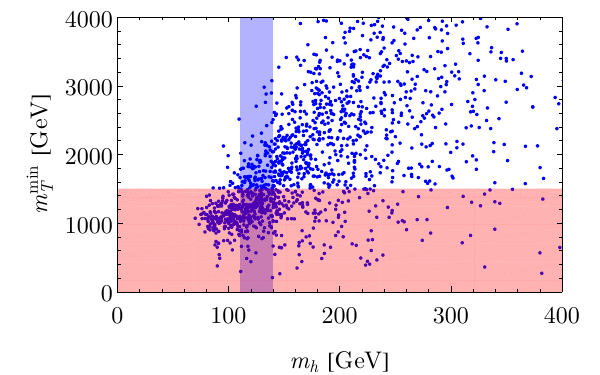}
    \caption{The lightest top--partner mass, $m_T^\textrm{min}$, versus the Higgs mass $m_h$. The shaded blue region is highlighting the correct Higgs mass $m_h \in (125 \pm 15)$ GeV, whereas the shaded red region shows current experimental limits on the lightest top partner, $m_T \gtrsim 1500$ GeV (\cite{CMS:2022fck,ATLAS:2018ziw,ATLAS:2022hnn,CMS:2019eqb,ATLAS:2022ozf}).}
    \label{fig:LightestTopPartner}
\end{figure}

\begin{figure}
    \centering
    \includegraphics[width=0.45\textwidth]{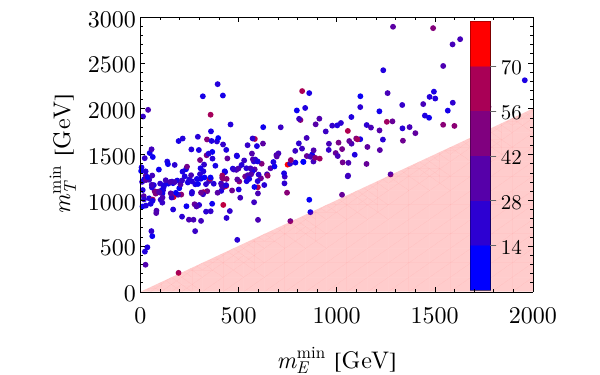}
    \caption{$m_T^\textrm{min}$ versus $m_{E}^\textrm{min}$. The colour--code corresponds to the tuning $\Delta_{\text{BG}}$. No viable points are found within the red region which strictly enforces the phenomenological bound $m_T^\textrm{min} \geq m_{E}^\textrm{min}$.
    }
    \label{fig:topexoticTuning}
\end{figure}

The results for the lightest top--partner mass, $m_T^{\textrm{min}}$, plotted against the Higgs mass are shown in Fig.~\ref{fig:LightestTopPartner}. The current LHC limit $m_T \gtrsim 1500$ GeV (\cite{CMS:2022fck,ATLAS:2018ziw,ATLAS:2022hnn,CMS:2019eqb,ATLAS:2022ozf}) is indicated by the red region. Filtering for $m_h \in (125 \pm 15)$ GeV, the spectrum of the lightest exotic, $m_E^{\textrm{min}}$, can be seen in Fig.~\ref{fig:topexoticTuning} in correlation with the top--partner mass. We observe that the exotic is strictly lighter than the top partner, providing an attractive collider target.

\begin{figure}
    \centering
    \includegraphics[width=0.45\textwidth]{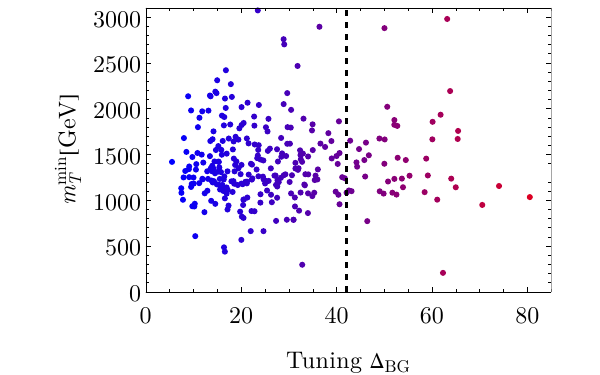}
    \caption{The lightest top--partner mass $m_T^\textrm{min}$ versus the amount of tuning $\Delta_{\text{BG}}$. The dashed black line shows the expected value for conventional minimal tuning $\Delta_{\text{min}} \sim 42$.}
    \label{fig:topTuning}
\end{figure}

The necessary fine--tuning to achieve the correct Higgs potential is assessed by employing the Barbieri--Giudice measure $\Delta_{\text{BG}}$~\cite{Barbieri:1987fn} i.e. the maximal sensitivity of observable $O$ to parameters $x_i$. We choose the Higgs mass and vacuum expectation value (vev) as observables to fully characterise the potential and take the maximum. We compare the obtained tuning, $\Delta_{\text{BG}}$, with the so--called minimal tuning~\cite{Panico:2012uw} $\Delta_{\text{min}} = f^2/v^2$, which is the minimal tuning in composite Higgs models that do not feature a double--tuning problem. In Fig.~\ref{fig:topTuning} the tuning is plotted against the mass of the lightest top partner, while filtering for correct Higgs mass and vev, showing $\Delta_{\text{min}} \sim 42$ for $f=1600$\,GeV as a dashed black line. There is a clustering of points with $\Delta_{\text{BG}} \sim 10 - 20$, which is comparable to minimal--tuning Composite Higgs models, but for the much lower (and phenomenologically problematic) scale of $f=800$ GeV for the latter. Furthermore, we see that heavy top partners are rather uncorrelated with the amount of tuning. 

\textbf{Phenomenology.}
Both $SO(11)/SO(10)$ and $SU(6)/SU(5)$ feature a global baryon symmetry. This property stems from the incomplete filling of elementary fermions into $G$ multiplets in the framework of partial compositeness as opposed to 4D GUTs in which the perfect filling of fermion representations break baryon number (see e.g.~\cite{Hosotani:2015hoa,Angelescu:2021nbp}). As a consequence, the proton remains stable, while the exotic fermions get charged with unusual baryon number $B=2/3$.  

Due to its peculiar baryon number, the exotic possesses unusual decay channels. In general, since it is lighter than both pNGB and vector LQs, the leading branching ratio corresponds to a three--body decay proceeding through an off--shell scalar or vector LQ. By imposing baryon number and electromagnetic charge conservation, the two possible 3--body decay channels are $\omega \to t b \tau^-$ and $\omega \to b b \nu$, where we expect decays to the more elementary 1st and 2nd generation of SM fermions to be suppressed. Moreover, since all three generations of lepton doublets are expected to be elementary, we envisage the $\omega \to b b \nu$ decay width to be suppressed by at least $m_\tau^2 / m_{top}^2$ with respect to $\omega \to t b \tau^-$. Therefore, we can safely take ${\rm BR} (\omega \to t b \tau^-)=1$.

The main production mechanism of the exotics at the LHC is through QCD pair production, $pp \to \omega \bar{\omega}$, leading to a peculiar $t \bar{t} b \bar{b} \tau^+ \tau^-$ final state. To the best of our knowledge, there is no dedicated search of the LHC collaborations for such a process, with traditional top and bottom partner searches not applying due to the different decay topologies, which is why the exotic can be much lighter than the conventional $B=1/3$ top partners. Nevertheless, from Fig.~\ref{fig:topexoticTuning} we observe that it could also be as heavy as 1500\,GeV, while keeping the tuning small.

Another potential signature of CH models is a change to Higgs production and decay. We note that for $f=1600$\,GeV these are in general safely below experimental limits, with the potential exceptions of new contributions due to (light) exotics. 
Importantly, we find that the latter do not spoil the gluon-fusion cross section because the opposite--chirality partners $\Tilde{\psi}'$ are elementary and do not interact directly with the Higgs.

\textbf{Conclusions.}
In this article we proposed a novel mechanism for generating the Higgs potential at subleading order in the fermion contributions by using a remarkable property of group representations. In contrast to twin Higgs models~\cite{Chacko:2005pe} (see~\cite{Geller:2014kta,Barbieri:2015lqa,Low:2015nqa} for composite scenarios) the quadratic contribution is cancelled via colored partners, which however carry a different global charge, therefore resulting in a different phenomenology. The cancellation relies on the decomposition of the pseudoreal representation $\bf{R} \rightarrow \bf{C} \oplus \Bar{\bf{C}}$ under $H$.

We analysed the setup in a three--site model showing a large reduction in fine--tuning to the $\sim 10\%$ level in comparison with the naive expectation which is at the percent level, or, in models that feature double--tuning, even worse. By virtue of the reduction in fine--tuning we can double the symmetry breaking scale to $f=1600$\,GeV and thus evade all top partner bounds, while keeping the fine--tuning comparable to minimal--tuning CH models with $f=800$ GeV. As a consequence of the unusual baryon number $B=2/3$ of the exotics, the expected signature of their decay is a six particle final state, which has not yet been targeted at the LHC. The search for signatures of natural models of electroweak symmetry breaking continues at the collider frontier.

\textbf{Acknowledgments}
We are grateful to Yi Chung, Lucia Masetti, Álvaro Pastor Gutiérrez, Aika Tada, and Stefan Tapprogge for useful discussion and comments.
\\
\appendix
\section{Proof of the Identity}\label{app1}
Here we present a proof of the following identity, involving the dressed spurions, necessary for the cancellation to work:
\begin{equation}
   {\rm Tr} \left[ (\Delta^{\Bar{\bf{C}}}_D)^\dagger \Delta^{\Bar{\bf{C}}}_D \right] = 
   {\rm Tr} \left[ (\Delta'^{\bf{C}}_D)^\dagger \Delta'^{\bf{C}}_D \right]
    \label{equ:trafoInvariants2}.
\end{equation}

The first step is to write the partial compositeness Lagrangian 
\begin{equation}
    \mathcal{L}_{\textrm{PC}} = \lambda\, \Bar{\psi} \Delta \,\mathcal{O}^{\bf{R}} + \lambda^\prime\, \Bar{\psi}^\prime \Delta^\prime \,\mathcal{O}^{\bf{R}} + \text{ h.c}\,.
    \label{equ:partialCompositenessGeneral},
\end{equation}
in a block--matrix form that makes the decomposition of $\bf R$ into $\bf C$ and $\bf \bar{C}$ apparent. In such a notation, the spurions, which are ${\rm dim}({\bf R}) \times {\rm dim}({\bf R})$ matrices, can be written as block matrices with four ${\rm dim}({\bf C}) \times {\rm dim}({\bf C})$ entries:
\begin{equation}
    \Delta = \begin{pmatrix}
        \delta & 0 \\ 0 & 0
    \end{pmatrix}, \quad \Delta^\prime = \begin{pmatrix}
        0 & 0 \\ 0 & \delta
    \end{pmatrix},
\end{equation}
with $\delta$ a diagonal matrix (with entries equal to 1 or 0 on the diagonal). The relation $\Delta_{11} = \Delta^\prime_{22} \equiv \delta$ follows from the fact that $\psi$ is the conjugate of $\psi^\prime$. After decomposing $\mathcal{O}^{\bf{R}}$ into $\mathcal{O}^{\bf{C}}$ and $\mathcal{O}^{\Bar{\bf{C}}}$, Eq.~\eqref{equ:partialCompositenessGeneral} becomes in this basis:
\begin{equation}
    \mathcal{L}_{\textrm{PC}} = \begin{pmatrix}
        \lambda \Bar{\psi} & \lambda^\prime \Bar{\psi^\prime}
    \end{pmatrix}
    \begin{pmatrix}
        \delta & 0 \\ 0 & \delta
    \end{pmatrix}
    \begin{pmatrix}
        \mathcal{O}^{\bf{C}} \\ \mathcal{O}^{\Bar{\bf{C}}}
    \end{pmatrix}  + \text{ h.c.}\,,
\end{equation}
where we truncated $\psi^{(\prime)}$ to span the relevant ${\rm dim}({\bf C})$ entries.
Performing an analogous decomposition to the Goldstone matrix $U$, we can compute the dressed spurions as:
\begin{equation}
\begin{pmatrix}
        \delta & 0 \\ 0 & \delta
        \end{pmatrix} 
        \begin{pmatrix}
        U_{11} & U_{12} \\ U_{21} & U_{22}
    \end{pmatrix}
    = \begin{pmatrix}
        \delta\, U_{11} & \delta\, U_{12} \\ \delta\, U_{21} & \delta\, U_{22}
    \end{pmatrix} \equiv \begin{pmatrix}
       \Delta^{\bf{C}}_D  & \Delta^{\Bar{\bf{C}}}_D \\ \Delta'^{\bf{C}}_D & \Delta'^{\Bar{\bf{C}}}_D
    \end{pmatrix}.
\end{equation}

The second step consists in deriving certain properties of the Goldstone matrix in the $\bf R$ representation of $G$. Since $\bf R$ is a pseudoreal representation, group theory tells us that for any $G$ transformation in the $\bf R$ representation, $ g = \exp{\left(i \, x_A T_{\bf R}^A\right)}$, with $A$ spanning the full set of $G$ generators, there exists an antisymmetric matrix $S = - S^T$, which can be chosen to be unitary, such that $S  g S^{-1} = g^*$. By inspecting transformations along the unbroken generators $T^a_{\bf R}$ and writing them in block matrix form,
\begin{equation}
\exp \left( i \, x_a T^a_{\bf R} \right) = \exp \left[ i \, x_a \begin{pmatrix}
T^a_{\bf C} & 0 \\ 0 & - \left(T^a_{\bf C}\right)^*
\end{pmatrix} \right],
\end{equation}
we deduce, without loss of generality, that that a suitable form for $S$ is 
\begin{equation}
    S = \begin{pmatrix}
        0 & 1 \\ -1 & 0
    \end{pmatrix}.
\end{equation}
Since the Goldstone matrix is itself a $G$ transformation, we have that $S  U S^{-1} = U^*$, from which it follows that
\begin{equation}
    U_{22} = U_{11}^*, \quad U_{21} = - U_{12}^*.
\end{equation}
Using this relation and the fact that $\delta$ is a diagonal matrix, Eq.~\eqref{equ:trafoInvariants2} follows straightforwardly.

\bibliography{ReferencesGHGUTPot}
\end{document}